\documentclass[conference,letterpaper]{IEEEtran}
\IEEEoverridecommandlockouts 
\usepackage[utf8]{inputenc} 
\usepackage[T1]{fontenc}
\usepackage{url}
\usepackage{amssymb}
\usepackage{ifthen}
\usepackage{cite}
\usepackage{pgfplots}
\pgfplotsset{compat=1.16}
\usepackage[cmex10]{amsmath} 

\newcommand\numeq[2]%
  {\stackrel{#1}{#2}}

\usepackage{epsfig,rotating,setspace,latexsym,amsmath,epsf,amssymb,amsfonts,bm,theorem,cite,algorithm,graphicx,epsf,authblk,epstopdf,color,algorithm,algpseudocode,bbm,subcaption}

\newtheorem{theorem}{Theorem}

\begin{document}

\title{Status Updating with Time Stamp Errors} 
\author{$\!\!\!\!\!\!$Md Nurul Absar Siddiky$ \qquad~~~$Ahmed Arafa\\Department of Electrical and Computer Engineering\\ University of North Carolina at Charlotte, NC 28223\\
$~~~~$\emph{msiddiky@charlotte.edu}$\qquad$\emph{aarafa@charlotte.edu}
\thanks{This work was supported by the U.S. National Science Foundation under Grants CNS 21-14537 and ECCS 21-46099.}}

\maketitle

\begin{abstract}
A status updating system is considered in which multiple processes are sampled and transmitted through a shared channel. Each process has its dedicated server that processes its samples before time stamping them for transmission. Time stamps, however, are prone to errors, and hence the status updates received may not be credible. Our setting models the time stamp error rate as a function of the servers' busy times. Hence, to reduce errors and enhance credibility, servers need to process samples on a relatively prolonged schedule. This, however, deteriorates timeliness, which is captured through the age of information (AoI) metric. An optimization problem is formulated whose goal to characterize the optimal processes' schedule and sampling instances to achieve the optimal trade-off between timeliness and credibility. The problem is first solved for a single process setting, where it is shown that a \emph{threshold-based sleep-wake schedule} is optimal, in which the server wakes up and is allowed to process newly incoming samples only if the AoI surpasses a certain threshold that depends on the required timeliness-credibility trade-off. Such insights are then extended to the multi-process setting, where two main scheduling and sleep-wake policies, namely round-robin scheduling with threshold-waiting and asymmetric scheduling with zero-waiting, are introduced and analyzed.
\end{abstract}

\section{Introduction}

\begin{figure}[t]
\centering
\includegraphics[scale=.575]{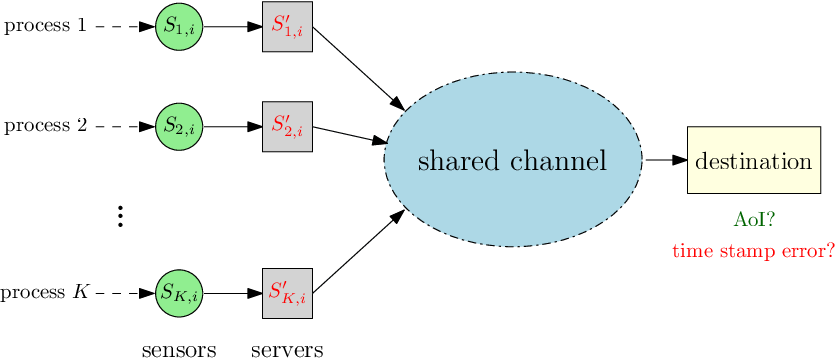}
\caption{System model: process $k$'s $i$th sample arrives at time $S_{k,i}$, yet is time-stamped as $S_{k,i}^\prime$ by its server.}
\label{fig_system}
\vspace{-.2in}
\end{figure}

Several current and emerging applications in communications, networking and control require timely information processing and transfer in order to accurately achieve their goals. This has led to the emergence of the age of information (AoI) metric, which assesses data freshness data at the destinations \cite{yates2021age}, and is defined as the difference between the current time and the time stamp of the latest received data \cite{kaul2012real}. In time-sensitive applications, it is crucial to measure the AoI accurately in order to take timely decisions. However, when time stamps are erroneous, the AoI value becomes unreliable, and the credibility of the decision-making process becomes questionable. In this work, we introduce the notion of {\it timeliness-credibility} trade-off through modeling analyzing the effects of time stamp errors on AoI optimization in a system where multiple processes are monitored through a shared communication channel, see Fig.~\ref{fig_system}.

Optimizing AoI, or maximizing timeliness and freshness of data, has been considered in a plethora of works in the literature. The pioneering work in queuing networks \cite{kaul2012real} and what follows in that line of research have shown that AoI-optimal policies are neither throughput-optimal (high server utilization) nor delay-optimal (low server utilization). Rather, AoI aims at balancing server utilization to deliver fresh data. Other lines of research to which these ideas are extended include, e.g., energy harvesting communications \cite{9516691}, federated learning \cite{yang2020age}, gossip networks \cite{yates2021age_gossip}, data trading \cite{9589847}, coding \cite{sac2018age}, internet-of-things (IoT) networks \cite{abd2019role}, random access networks \cite{zhao2024optimizing}, edge computing \cite{he2024age}, and privacy-preserving systems \cite{arafa2023private}.

A notable challenge to achieve accurate AoI is the issue of \textit{timestomping} \cite{minnaard2014timestomping}, where time stamps on data packets are intentionally falsified, typically as part of an adversarial attack. This manipulation can occur through several mechanisms, including falsifying time stamps to make stale data appear fresh, introducing network delays, or due to natural factors such as sensor malfunctions or synchronization issues. These time stamp errors lead to inaccurate AoI calculations, which in turn result in misleading assessment of data freshness. This performance degradation is particularly concerning in energy-constrained IoT systems, remote monitoring, and decentralized networks, where accurate updates are critical. In \cite{kaswan2024timestomping}, the effects of adversarial time stamp manipulation in gossip networks have been studied, demonstrating that even a compromised node in fully connected networks can significantly increase AoI and worsen how it scales with the size of the network. 

Although time stamp manipulation has been explored in gossip networks, there is a lack of research on how time stamp inaccuracies affect AoI in conventional update systems. To address this gap, our work investigates the {\it optimization of both AoI and time stamp accuracy,} with the goal of enhancing system reliability and timeliness. Minimizing AoI alone is insufficient when time stamp errors are present, as these errors can lead to poor decision making in critical applications such as remote sensing and energy-constrained systems. Therefore, we propose integrating time stamp error management into AoI optimization to provide more accurate and reliable decision-making and system performance.

Specifically, we consider a system in which time stamps from multiple processes are sent through a shared channel towards a destination. Each process has a dedicated server to process its samples and assign them time stamps before sending them on the channel. A server introduces time stamp errors with a rate that depends on its busy time. That is, to reduce the errors, a server needs to {\it sleep} for a while. This, in turn increases the AoI and reduces timeliness. Hence, a trad-off arises between minimizing AoI and minimizing time stamp errors. We introduce an optimization problem to characterize the optimal trade-off by optimizing the sleep-wake schedules of the servers. We first solve the problem for the single process setting. Towards that end, we show that the optimal sleep-wake schedule has a {\it threshold structure:} the server wakes up only if the AoI surpasses a certain threshold that depends on the target credibility of time stamps. We then build on these insights and present two main scheduling policies for the multiple-source setting: round-robin scheduling with threshold-waiting, and asymmetric scheduling with zero-waiting. We analyze and compare the performances of both policies and show that the optimal choice between them highly depends on the system parameters and the target time stamp credibility.

\section{System Model and Problem Formulation} \label{sec_sys}

We consider a status update system composed of $K$ sensors and $K$ servers, as shown in Fig.~\ref{fig_system}. Sensor $k$ receives samples from a $\lambda_k$-Poisson process, and passes them to server $k$ for time-stamping and transmission. Transmissions go through a shared communication channel that adds a random delay and can only be utilized by one server at a time.

Let $\pi$ denote the transmission schedule. This schedule also defines a sleep-wake policy for servers (and sensors); to avoid samples becoming stale, sensor $k$ does not acquire new samples and goes into sleep mode unless it is scheduled to transmit according to $\pi$. More precisely, when sensor $k$'s turn comes up for the $i$th time, it may choose to continue to sleep (or wait) for $W_{k,i}$ extra time units, after which it wakes up and becomes ready to receive new samples. Then, it receives its $i$th sample after $X_{k,i}$ time units. Note that $X_{k,i}$'s are independent and identically distributed (i.i.d.). $\sim\text{exp}(\lambda_k)$ across process $k$ samples. We denote by $\{W_{k,i}\}$ the servers' waiting policy.

Now, let $S_{k,i}$ denote the arrival time of the $i$th sample of the $k$th process. Such sample gets served immediately upon arrival, and reaches the destination at time
\begin{align}
D_{k,i}=S_{k,i}+Y_{k,i},
\end{align}
where $Y_{k,i}$'s are i.i.d. across samples and processes, denoting channel busy times. We assume that the destination is {\it unaware} of the values of the channel busy times $\{Y_{k,i}\}$; only the sensor-server side is aware.

\begin{figure}[t]
\centering
\includegraphics[scale=.6]{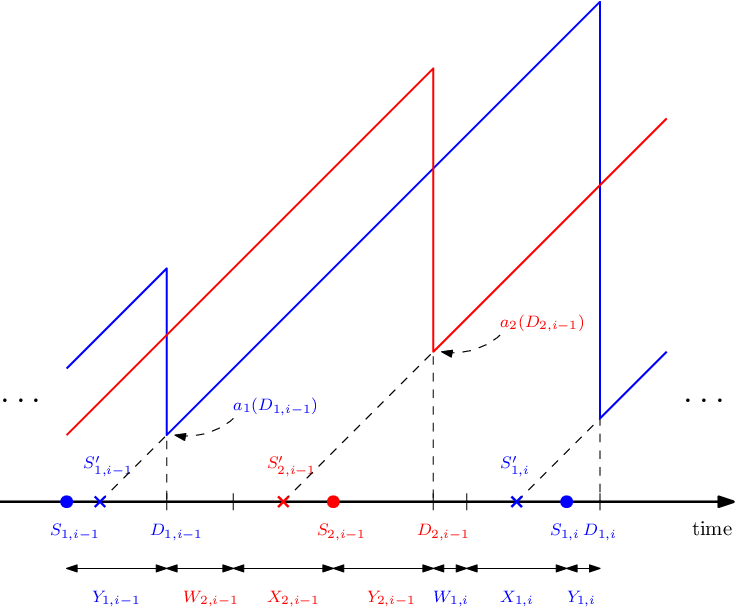}
\caption{Example AoI evolution at the destination for $K=2$ processes ($1$ in blue; $2$ in red). Filled circles denote true time stamps and crosses denote received (erroneous) time stamps.}
\label{fig_aoi-evol}
\vspace{-.2in}
\end{figure}

Servers may introduce time stamp errors, in which the {\it received} time stamp of the process $k$'s $i$th sample is given by $S^\prime_{k,i}$ as opposed to the true time stamp $S_{k,i}$. Errors occur at a rate that depends on the sleep-wake schedule of the sensors as we explain below. Statistically, we assume that $S^\prime_{k,i}$ and $S_{k,i}$ are related as follows:
\begin{align}
\mathrm{E}\left[S^\prime_{k,i}|S_{k,i},D_{k,i}\right]&=S_{k,i}, \label{eq_avg-stmp} \\
\mathrm{Var}\left(S^\prime_{k,i}|S_{k,i},D_{k,i}\right)&=h_k\left(S_{k,i}-X_{k,i}-S_{k,i-1}\right), \label{eq_var-stmp}
\end{align}
where $\mathrm{E}[\cdot]$ and $\mathrm{Var}(\cdot)$ denote expectation and variance, respectively, and $h_k(\cdot)$ is some monotonically {\it decreasing convex} function. Hence, the server's introduced (and received) time stamp $S^\prime_{k,i}$ is unbiased from the true time stamp $S_{k,i}$, yet its variance is inversely proportional with the inter-sampling duration. Therefore, to reduce errors of a certain server, one needs to reduce its sampling rate. The rationale is that {\it errors occur more often when servers do not get enough sleep time.} Such approach has been considered previously in, e.g., use-dependant channels \cite{ward2017use}. While our results are presented for general functions $h_k(\cdot)$, our experiments will be focusing on exponentially-decaying functions given by
\begin{align} \label{eq_h-exp}
h_k(x)=e^{-\alpha_kx},
\end{align}
for some parameter $\alpha_k\geq0$ denoting server $k$'s {\it recovery rate}. That is, higher values of $\alpha_k$ represent faster recovery, in which server $k$ introduces relatively less errors and can tolerate being awake for relatively longer periods of time, and vice versa.

We assess timeliness at the destination using AoI. For process $k$, the AoI at time $t$ is defined as
\begin{align}
a_k(t)=t-S^\prime_{k,i},\quad D_{k,i}\leq t<D_{k,i+1}.
\end{align}
An example of the AoI evolution at the destination for $K=2$ processes is shown in Fig.~\ref{fig_aoi-evol}. Note that due to timestamping errors, the AoI value seen at the destination may not represent the true value of the AoI. We define by an {\it epoch} the time elapsed in between two consecutive deliveries of samples from a specific process. Let us denote by
\begin{align} \label{eq_epoch-k-length}
L_{k,i}\triangleq D_{k,i}-D_{k,i-1}
\end{align}
the length of the $i$th epoch for process $k$. We are interested in the long-term average AoI for process $k$ defined by the area under its AoI curve. From Fig.~\ref{fig_aoi-evol}, such quantity is given by
\begin{align}\label{AoI_main}
\overline{\texttt{AoI}}_k\!=\!\limsup_{n\rightarrow\infty}\frac{\sum_{i=1}^n\!\mathrm{E}\left[a_k\left(D_{k,i-1}\right)L_{k,i}\right]+\frac{1}{2}\mathrm{E}\left[L_{k,i}^2\right]}{\sum_{i=1}^n\mathrm{E}\left[L_{k,i}\right]}.
\end{align}

As we can see from the above, the timeliness measured at the destination is {\it not always credible} due to time stamping errors. Therefore, one also needs to measure the long-term average error for a specific process when evaluating its timeliness. Such quantity is given by
\begin{align}\label{error_main}
\overline{\texttt{e}}_k=\limsup_{n\rightarrow\infty}\frac{1}{n}\sum_{i=1}^n\mathrm{E}\left[\left(S_{k,i}-S^\prime_{k,i}\right)^2\right].
\end{align}

Using \eqref{eq_avg-stmp} and \eqref{eq_var-stmp}, observe that one can reduce the value of $\overline{\texttt{e}}_k$ by increasing the inter-sampling duration. However, this may negatively impact timeliness. Hence, {\it a trade-off arises between timeliness and credibility.} Note that a schedule $\pi$ combined with a waiting policy $\{W_{k,i}\}$ completely characterize the values of $\overline{\texttt{AoI}}_k$ and $\overline{\texttt{e}}_k$ for all processes. Our main goal is to optimize a weighted average of timeliness and credibility. That is, to solve the following optimization problem:
\begin{align} \label{opt_main}
\min_{\pi,~\{W_{k,i}\geq0\}} \quad \sum_{k=1}^K\beta_k\overline{\texttt{AoI}}_k+\left(1-\beta_k\right)\overline{\texttt{e}}_k,
\end{align}
for some $\beta_k\in[0,1],~\forall k$.

We first solve the single-process version of the above problem in the next section. Then, we present solutions for the multi-process version in the following one.

\section{The Single Process Setting} \label{sec_sgl}

In the case of $K=1$ process, we drop the index $k$ from all the variables, and drop the schedule $\pi$ from the optimization problem in \eqref{opt_main}. That is, the only variable of the optimization problem in \eqref{opt_main} is now the waiting policy $\{W_i\}$. 

Towards characterizing the long-term average AoI, observe that the starting AoI of epoch $i$ is given by
\begin{align}
a\left(D_{i-1}\right)=&\begin{cases}Y_{i-1}-\left(S^\prime_{i-1}-S_{i-1}\right),\quad S^\prime_{i-1}\geq S_{i-1}\\
Y_{i-1}+S_{i-1}-S^\prime_{i-1},\quad S^\prime_{i-1}<S_{i-1}\end{cases} \nonumber \\
=&Y_{i-1}+S_{i-1}-S^\prime_{i-1}. \label{eq_strt-aoi}
\end{align}

Next, we focus on {\it stationary deterministic} waiting policies in which the waiting time in epoch $i$ is given by a deterministic function of the channel busy time in epoch $i-1$. That is,
\begin{align}
W_i\triangleq \omega\left(Y_{i-1}\right),
\end{align}
for some function $\omega(\cdot)$ to be optimized. Such waiting policy has been shown optimal in similar settings of AoI optimization \cite{arafa2023private}, in which the channel busy times are i.i.d.\footnote{Observe that the waiting policy is determined completely at the sensor-server side where full knowledge of the channel busy times $Y_i$'s is provided. Hence, the policy is implementable in our setting.} Such choice of waiting policies induces a stationary distribution across epochs. Specifically, the $i$th epoch length is now given by
\begin{align}
L_i=\omega\left(Y_{i-1}\right)+X_i+Y_i,
\end{align}
and the long-term average AoI now reduces to the following:
\begin{align}
\overline{\texttt{AoI}}=\frac{\mathrm{E}\left[a\left(D_{i-1}\right)L_i\right]+\frac{1}{2}\mathrm{E}\left[L_i^2\right]}{\mathrm{E}\left[L_i\right]}.
\end{align}

Now let us further analyze the term $\mathrm{E}\left[a\left(D_{i-1}\right)L_i\right]$ in the numerator above. Using \eqref{eq_avg-stmp} and \eqref{eq_var-stmp}, one can see that the time stamp error $S_{i-1}-S^\prime_{i-1}$ only depends on channel busy time $Y_{i-2}$, and is therefore independent from $L_i$. Further, one can show that its average value is equal to $0$ as follows:
\begin{align}
\mathrm{E}&\left[S_{i-1}-S^\prime_{i-1}\right]=\mathrm{E}\left[\mathrm{E}\left[S_{i-1}-S^\prime_{i-1}|S_{i-1},D_{i-1}\right]\right] \nonumber \\
&=\mathrm{E}\left[\mathrm{E}\left[S^\prime_{i-1}|S_{i-1},D_{i-1}\right]-\mathrm{E}\left[S^\prime_{i-1}|S_{i-1},D_{i-1}\right]\right]=0. 
\end{align}
Therefore, using \eqref{eq_strt-aoi} we get that
\begin{align}
\mathrm{E}\left[a\left(D_{i-1}\right)L_i\right]=\mathrm{E}\left[Y_{i-1}L_i\right].
\end{align}
In other words, the time stamp error, on average, does not affect the long-term average AoI viewed from the destination.\footnote{This, however, necessitates the addition of a credibility measure as in \eqref{error_main}.}

We now turn to the long-term average error. Since we have stationary distributions across epochs, we get
\begin{align}
\overline{\texttt{e}}=&\mathrm{E}\left[\left(S_i-S^\prime_i\right)^2\right] \nonumber \\
=&\mathrm{E}\left[\mathrm{E}\left[\left(S_i-S^\prime_i\right)^2|S_i,D_i\right]\right] \nonumber \\
=&\mathrm{E}\left[\mathrm{E}\left[\left(\mathrm{E}\left[S^\prime_i|S_i,D_i\right]-S^\prime_i\right)^2|S_i,D_i\right]\right] \nonumber \\
=&\mathrm{E}\left[\mathrm{Var}\left(S^\prime_i|S_i,D_i\right)\right] \nonumber \\
=&\mathrm{E}\left[h\left(S_i-X_i-S_{i-1}\right)\right] \nonumber \\
=&\mathrm{E}\left[h\left(Y_{i-1}+\omega\left(Y_{i-1}\right)\right)\right].
\end{align}

The optimization problem is now given by
\begin{align}
\min_{\omega(\cdot)\geq0} \quad \overline{\texttt{AoI}} + \frac{1-\beta}{\beta} \overline{\texttt{e}},
\end{align}
which can be equivalently represented as 
\begin{align} \label{opt_sgl-tau}
\min_{\omega(\cdot)\geq0} \quad \overline{\texttt{AoI}},\quad \mbox{s.t.} \quad \overline{\texttt{e}}\leq\tau,
\end{align}
for some $\tau\geq0$ \cite{boyd2004convex}. We call the constraint in \eqref{opt_sgl-tau} the {\it credibility constraint.} We focus on analyzing the second formulation in \eqref{opt_sgl-tau} in the remainder of this section. Specifically, we follow Dinkelbach's approach \cite{dinkelbach1967nonlinear} to transform the problem into the following auxilliary one:
\begin{align} \label{opt_sgl-aux}
p(\theta)\triangleq\min_{\omega(\cdot)\geq0} \quad &\mathrm{E}\left[\left(Y_{i-1}-\theta\right)\left(\omega\left(Y_{i-1}\right)+X_i+Y_i\right)\right] \nonumber \\
&+\frac{1}{2}\mathrm{E}\left[\left(\omega\left(Y_{i-1}\right)+X_i+Y_i\right)^2\right] \nonumber \\
\mbox{s.t.} \quad &\mathrm{E}\left[h\left(Y_{i-1}+\omega\left(Y_{i-1}\right)\right)\right]\leq\tau,
\end{align}
for some $\theta\in\mathbb{R}$. The optimal solution of problem \eqref{opt_sgl-tau} is now given by the unique $\theta^*$ that solves $p(\theta^*)=0$, which can be found by, e.g., a bisection search.

The objective function of problem \eqref{opt_sgl-aux} can be further simplified as follows:
\begin{align}
&\mathrm{E}\left[\left(Y_{i-1}-\theta\right)\omega\left(Y_{i-1}\right)\right]+\left(\mu_Y-\theta\right)\left(\frac{1}{\lambda}+\mu_Y\right) \nonumber \\
&+\frac{1}{2}\mathrm{E}\left[\omega\left(Y_{i-1}\right)^2\right]+\mathrm{E}\left[\omega\left(Y_{i-1}\right)\right]\left(\frac{1}{\lambda}+\mu_Y\right) \nonumber \\
&+\frac{1}{\lambda^2}+\frac{1}{\lambda}\mu_Y+\frac{1}{2}\mu_{Y^2},
\end{align}
where $\mu_Y$ and $\mu_{Y^2}$ denote the first and second moments of $Y_i$, respectively. We now introduce the following Lagrangian for problem \eqref{opt_sgl-aux}:
\begin{align}
\mathcal{L}=&\int\left(\left(y-\theta+\frac{1}{\lambda}+\mu_Y\right)\omega(y)+\frac{1}{2}\omega(y)^2\right)f_Y(y)dy \nonumber \\
&+\left(\mu_Y-\theta\right)\left(\frac{1}{\lambda}+\mu_Y\right)+\frac{1}{\lambda^2}+\frac{1}{\lambda}\mu_Y+\frac{1}{2}\mu_{Y^2} \nonumber \\
&+\gamma\left(\int h\left(y+\omega(y)\right)f_Y(y)dy-\tau\right)-\int\eta(y)\omega(y)dy,
\end{align}
where $f_Y(y)$ denotes the distribution of $Y_i$, whereas $\gamma$ and $\eta(y)$ are Lagrange multipliers. Taking the functional derivative of the above with respect to $\omega(y)$, equating to $0$, and rearranging, we get
\begin{align} \label{eq_kkt4}
y+\omega(y)+\gamma h^\prime\left(y+\omega(y)\right)=\theta-\frac{1}{\lambda}-\mu_Y+\frac{\eta(y)}{f_Y(y)},
\end{align}
where $h^\prime(\cdot)$ denotes the derivative of $h(\cdot)$. Now observe that since $h(\cdot)$ is convex, and $\gamma\geq0$, the function 
\begin{align}
H_\gamma(x)\triangleq x+\gamma h^\prime(x)
\end{align}
is monotonically increasing. Hence, by \eqref{eq_kkt4} we have
\begin{align}
\omega(y)=H_\gamma^{-1}\left(\theta-\frac{1}{\lambda}-\mu_Y+\frac{\eta(y)}{f_Y(y)}\right)-y.
\end{align}
By complementary slackness \cite{boyd2004convex}, we get $\eta(y)=0$ if $\omega(y)>0$, in which case $H_\gamma^{-1}\left(\theta-\frac{1}{\lambda}-\mu_Y\right)>y$. On the other hand, if $H_\gamma^{-1}\left(\theta-\frac{1}{\lambda}-\mu_Y\right)<y$, then we must have $\eta(y)>0$ so as to increase the argument inside $H_\gamma^{-1}$ and make $\omega(y)$ non-negative. This means, again by complementary slackness, that $\omega(y)=0$. Combining the arguments, we finally have the optimal waiting policy that solves problem \eqref{opt_sgl-aux} given by
\begin{align} \label{eq_omega-star}
\omega^*(y)=\left[H_{\gamma^*}^{-1}\left(\theta-\frac{1}{\lambda}-\mu_Y\right)-y\right]^+
\end{align}
where $[\cdot]^+\triangleq\max(\cdot,0)$ and $\gamma^*$ denotes the optimal Lagrange multiplier associated with the credibility constraint.

The result above shows that the optimal waiting policy has a threshold structure; as long as the starting AoI of an epoch is below a certain threshold, given by $H_{\gamma^*}^{-1}\left(\theta-\frac{1}{\lambda}-\mu_Y\right)$, the sensor should continue in sleeping mode until the AoI surpasses that threshold, and then wake up. The threshold, however, remains partially unknown unless we can evaluate $\gamma^*$. We do so indirectly as follows. First, let us assume that $\gamma^*=0$. In this case, the threshold is simply given by
\begin{align}
H_0^{-1}\left(\theta-\frac{1}{\lambda}-\mu_Y\right)=\theta-\frac{1}{\lambda}-\mu_Y.
\end{align}
We now use the above to check if the credibility constraint is satisfied. If it is not, then it must be that $\gamma^*>0$, which means by complementary slackness that the credibility constraint is satisfied with equality. In this case, all we need to do is to find some threshold value $\xi$ that solves
\begin{align} \label{eq_thrshld-equl}
\mathrm{E}\left[h\left(Y_{i-1}+\left[\xi-Y_{i-1}\right]^+\right)\right]=\tau,
\end{align}
which can be evaluated by, e.g., a bisection search since $h(\cdot)$ is monotonically decreasing.

The approach above provides the optimal solution of problem \eqref{opt_sgl-aux}, i.e., it evaluates $p(\theta)$. The final step to link all this back to problem \eqref{opt_sgl-tau} is to find the optimal value $\theta^*$ that solves $p(\theta^*)=0$. We have now proven the following theorem that summarizes the theoretical results in this section:

\begin{theorem}
The optimal solution of problem \eqref{opt_sgl-tau} is given by a threshold-waiting policy $\omega^*(\cdot)=[\xi^*-\cdot]^+$. The threshold $\xi^*$ is given by $\theta^*-\frac{1}{\lambda}-\mu_Y$, provided that the credibility constraint is satisfied. Otherwise, it is given by the solution of \eqref{eq_thrshld-equl}. The value of $\theta^*$ is such that $p(\theta^*)=0$ in \eqref{opt_sgl-aux}.
\end{theorem}

\section{The Multi-Process Setting} \label{sec_mlti}

We now use the results developed for the single process setting to present solutions for problem \eqref{opt_main} in the case of $K\geq2$ processes. We first note that finding the jointly optimal scheduling and waiting policies is highly nontrivial. One reason behind this is that the functions governing the time-stamp errors, $h_k(\cdot)$'s, can vary from one server/process to another. For example, let us consider the exponentially decaying model in \eqref{eq_h-exp} for $K=2$. If $\alpha_1>\alpha_2$, then server 1 {\it recovers faster} than server 2. Hence, it could be optimal to schedule process 1 {\it more often} than process 2 so as to allow a sufficient time for server 2 to recover and reduce errors. Thus, the often studied round-robin schedule (or maximum-age-first) in the AoI literature \cite{Banawan_journal_2023} may not be optimal in our setting. This, in addition to the fact that we need to evaluate optimal waiting times renders the problem challenging.

To alleviate this hurdle, in this preliminary work on this problem we aim at developing policies that optimize the scheduling policy or the waiting policy {\it individually}, as opposed to jointly, and compare their performances against each other. We focus on two specific kinds of policies that we discuss next. We note that the detailed expressions of the resulting AoI's and time stamp errors from such policies are omitted due to space limits, and are demonstrated in the experimental results in Section~\ref{sec_num}.

\subsection{Round-Robin Scheduling with Threshold-Waiting}

The first policy that we discuss considers a round-robin (RR) schedule, in which process $1$ is scheduled for sampling, followed $2$, all the way to $K$, and then the schedule repeats. We denote this schedule by $\pi_{RR}$.

As for the waiting policy, and given the results of the single process setting, we combine RR scheduling with a threshold-waiting policy, which is illustrated as follows. Instead of waiting prior to each process sampling, we only wait once before scheduling process $1$. This is then followed by the scheduled RR transmissions. We note that such approach has been shown optimal in the relatively similar setting of \cite{Banawan_journal_2023}. Hence, let us focus on process $1$'s epoch. The waiting time at the beginning of epoch $i$ is now given by the following function of epoch $i-1$'s sum service times:
\begin{align} \label{eq_omega-RR}
\omega\left(\sum_{k=1}^KY_{k,i-1}\right)=\left[\xi-\sum_{k=1}^KY_{k,i-1}\right]^+,
\end{align}
for some threshold $\xi$ to be optimized. For simplicity of presentation, we refer to the above expression by $\omega_i^{RR}$.

Now observe that for the $k$th process, we get from \eqref{eq_epoch-k-length} that
\begin{align}
L_{k,i}\!=\!\!\sum_{s=k+1}^KX_{s,i-1}+Y_{s,i-1}+\omega_i^{RR}+\sum_{s=1}^kX_{s,i}+Y_{s,i}.
\end{align}
The above expression, together with the definition of $\omega_i^{RR}$ in \eqref{eq_omega-RR}, shows that the location of the waiting time could possibly lead to different distributions of epochs across different processes. However, we note that the {\it sums} of their corresponding AoI's and time stamp errors would still be the same, denoted by $\overline {\texttt{AoI}}_k (\pi_{RR})$ and $\overline {\texttt{e}}_k (\pi_{RR})$, respectively. We omit such details due to space limits. Based on this, we focus on the case in which all $\beta_k$'s are equal in this work. This implies that the relationship between AoI and the time-stamp error remains consistent across all processes, leading to a uniform stationary behavior in the system's performance.\footnote{The general case in which $\beta_k$'s are not equal can be solved by optimizing the {\it location} of the waiting time, and is to be analyzed in future work.}

Using the above expression, we simplify equations \eqref{AoI_main} and \eqref{error_main} and specialize them to the case of RR scheduling with threshold-waiting. The resulting expressions are used for specific service time distribution to get closed-form expressions for $\overline {\texttt{AoI}}_k (\pi_{RR})$ and $\overline {\texttt{e}}_k (\pi_{RR})$ in terms of the threshold $\xi$, which can then be found by, e.g., line search algorithms.

\subsection{Asymmetric Scheduling with Zero-Waiting}

The second policy under consideration is an asymmetric scheduling (AS) policy, indicated by \(\pi_{AS}\). In here, process 1 is scheduled for \(m_1\) sampling and transmission trials, followed by process 2 for \(m_2\) trials, and so on until process \(K\) completes its \(m_k\) trials, after which the schedule repeats. In this asymmetric schedule, different from $\pi_{RR}$, we do not consider waiting prior to transmission.

Using $\pi_{AS}$, an epoch for any process $k$ includes the same number of (possibly different) transmissions from every other process. Since the waiting time is $0$ and all random variables are i.i.d., it follows that the system is stationary and all epochs' distributions are the same. We denote the corresponding AoI and time stamp error for process $k$ by $\overline {\texttt{AoI}}_k (\pi_{AS})$ and $\overline {\texttt{e}}_k (\pi_{AS})$, respectively.

Let us denote by $X_{k,i}^{(j)}$ and $Y_{k,i}^{(j)}$ the $j$th inter-arrival and service times of the $k$th process in the $i$th epoch, respectively, where the epoch index is counted with respect to process $1$ without loss of generality. Therefore, we have
\begin{align}
\!\!L_{k,i}\!=\!\!\sum_{s=k+1}^K\sum_{j=1}^{m_s}X_{s,i-1}^{(j)}+Y_{s,i-1}^{(j)}\!+\!\sum_{s=1}^k\sum_{j=1}^{m_s}X_{s,i}^{(j)}+Y_{s,i}^{(j)}.
\end{align}
Based on the above expression, we specialize the equations in \eqref{AoI_main} and \eqref{error_main} and derive the AoI and time stamp error expressions for AS scheduling with zero waiting. Given a service time distribution, one can then find the optimal selection of the number of trials $m_k$ for process $k$.

\section{Numerical Results} \label{sec_num}

In this section, we present some numerical results to further illustrate the theoretical analysis of this paper. We focus on showing the AoI vs.~ time stamp error trade-off under different system settings. From the optimization problem in \eqref{opt_main}, such a trade-off can be characterized by varying the values of $\beta_k$'s. In our simulations, and in agreement with our theoretical results, we focus on the case in which $\beta_k=\beta,~\forall k$. The service time distribution is $\sim\text{exp}(1/\mu_Y)$. 

\begin{figure}[t]
    \centering   \includegraphics[width=1\linewidth]{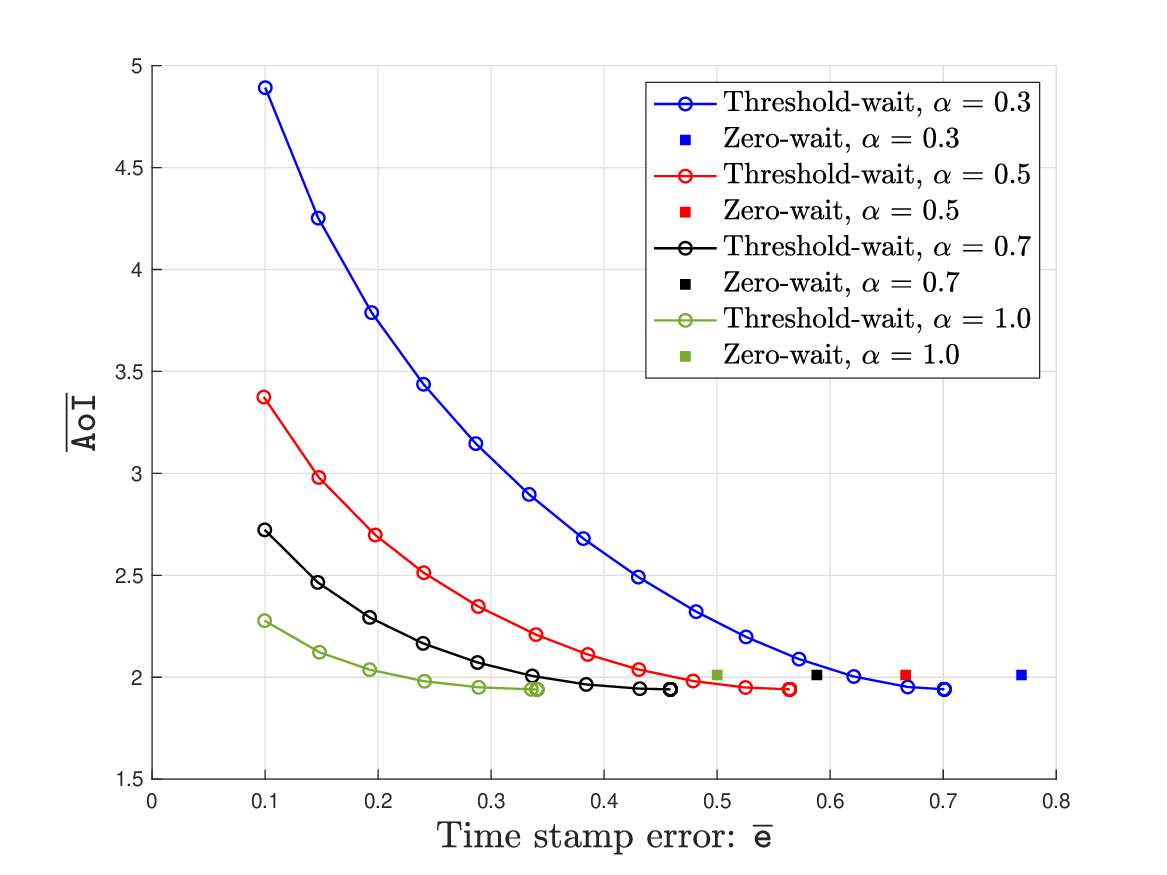}
    \caption{Single process AoI vs.~time-stamp error for different recovery rate \(\alpha\) values.}
    \label{fig_Single_Source_alpha}
    \vspace{-.1in}
\end{figure}

We first present results for the single process setting. In Fig.~\ref{fig_Single_Source_alpha}, we vary the value of $\beta\in[0,1]$ to show how the AoI behaves with respect to time stamp error. We set $\lambda=9$ and $\mu_Y=1$. Clearly, the higher the value of $\beta$ the better the AoI and the worse the error, and vice versa. Moreover, as the recovery rate $\alpha$ increases, the trade-off behaves better: for a relatively higher value of $\alpha$, one can achieve lower errors for the same AoI values. As a baseline, we show the single AoI-error pair achieved by the zero-wait policy. It is clear from the figure that the optimal threshold-wait policy outperforms zero-wait in terms of both AoI and error for relatively higher values of $\beta$.

\begin{figure}[t]
    \centering
    \includegraphics[width=1\linewidth]{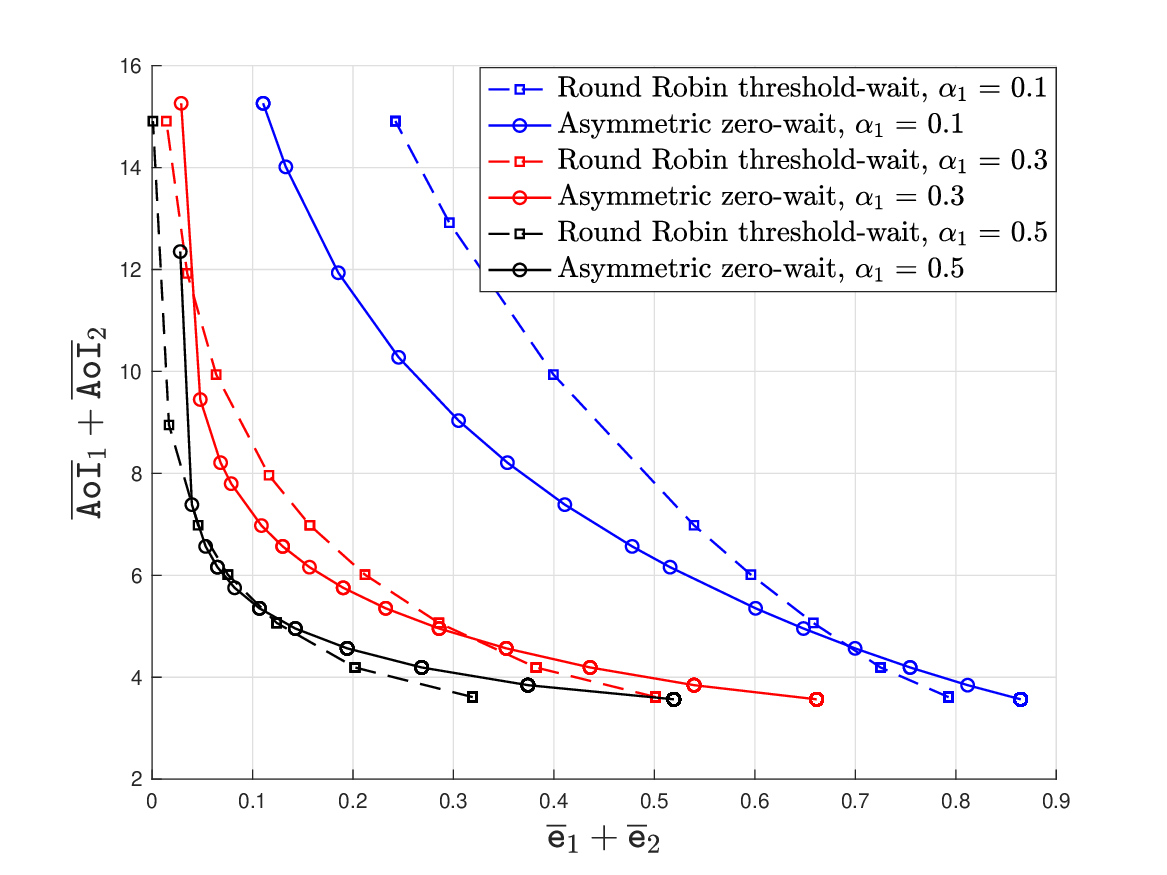}
    \caption{Two processes sum AoI vs.~sum time stamp error for different process 1 recovery rate \(\alpha_1\) values.}
    \label{fig_DoubleSource_alpha1}
    \vspace{-.1in}
\end{figure}

Next, we compare the behavior of $\pi_{RR}$ and $\pi_{AS}$ for $K=2$ processes. We set \(\lambda_1 = \lambda_2 = 6\), \(\alpha_2 = 50\), and \(\mu = 1.5\), and plot the sum AoI vs.~sum time stamp error (by varying $\beta$) in Fig.~\ref{fig_DoubleSource_alpha1}. The results generally show that the behavior of $\pi_{RR}$ relative to $\pi_{AS}$ depends on the value of $\alpha_1$ and $\beta$. For instance, for smaller $\alpha_1$ and smaller $\beta$, $\pi_{AS}$ is favored upon $\pi_{RR}$. While for larger $\alpha_1$ and larger $\beta$ the situation is reversed. For intermediate values no specific policy dominates the other. This shows that the choice of the scheduling and waiting policy for this problem is highly dependent on the system dynamics, especially the servers' recovery rates.

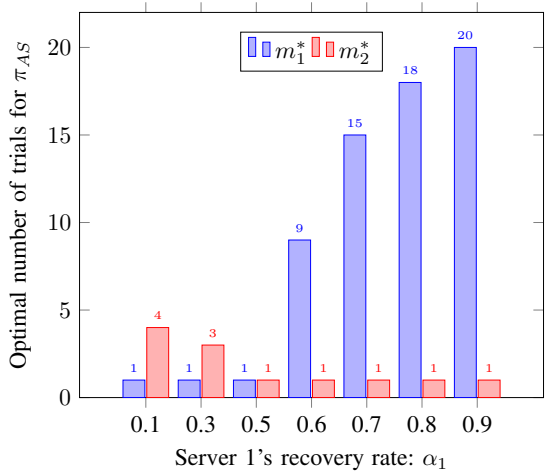
\begin{figure}[t]
\centering
\begin{tikzpicture}[scale=.9]
\begin{axis}[
    xlabel={Server 1's recovery rate: \(\alpha_1\)},
    ylabel={Optimal number of trials for $\pi_{AS}$},
    symbolic x coords={0.1, 0.3, 0.5, 0.6, 0.7, 0.8, 0.9},
    xtick=data,
    ybar=5pt, 
    bar width=9.2pt, 
    legend style={
        at={(0.5,0.95)},
        anchor=north,
        legend columns=-1
    },
    ymin=0,
    enlarge x limits=0.2, 
    nodes near coords, 
    nodes near coords align={vertical}, 
    every node near coord/.append style={font=\tiny} 
]

\addplot+[bar shift=-5.0pt] 
    coordinates {(0.1,1) (0.3,1) (0.5,1) (0.6,9) (0.7,15) (0.8,18) (0.9,20)};
\addplot+[bar shift=5.0pt] 
    coordinates {(0.1,4) (0.3,3) (0.5,1) (0.6,1) (0.7,1) (0.8,1) (0.9,1)};

\legend{\(m_1^*\),\(m_2^*\)}
\end{axis}
\end{tikzpicture}
\caption{Optimal AS policy behavior: $(m_1^*,m_2^*)$ vs.~$\alpha_1$.}
\label{fig_Variation of optimal transmission counts}
\vspace{-.1in}
\end{figure}

Finally, the column chart in Fig.~\ref{fig_Variation of optimal transmission counts} illustrates the optimal number of trials for $\pi_{AS}$, \(m_1^*\) and \(m_2^*\), vs.~different values of server 1's recovery rate \(\alpha_1\). Here, we set \(\alpha_2 = 0.5\), \(\lambda_1 = \lambda_2 = 90\), \(\mu = 50\), and \(\beta = 0.5\). As \(\alpha_1\) increases, server $1$ recovers relatively faster than server $2$, and therefore \(m_1^*\) increases while $m_2^*$ decreases as seen in the figure. This trend underscores the importance of tailoring transmission strategies to specific server recovery rates to optimize timeliness and credibility.

\section{Conclusion}

The impact of timestamp errors on the credibility of AoI in status updating systems has been investigated. Samples from multiple processes are acquired by sensors and then processed by servers to be sent through a shared channel. Through modeling time stamp error rates as a function of the servers' busy times, a trade-off has been introduced: allowing servers more sleeping time, and hence more time to recover, decreases time stamp errors, but increases AoI. An optimization problem has been formulated to characterize the optimal timeliness-credibility trade-off by designing scheduling and server sleep-wake policies. Solutions have been presented first for the single process setting, in which the optimal sleep-wake policy has been shown to have a threshold structure. For the multi-process setting, round-robin (symmetric) and asymmetric scheduling have been studied. Our results show that server recovery rates can highly affect the timeliness-credibility trade-off curves, and that scheduling policies should be chosen based on the system parameters, including processes' sampling rates, channel service rate and relative recovery rates among the servers.

\bibliographystyle{unsrt}
\bibliography{IEEEabrv,NurLibrary}

\begin{thebibliography}{10}

\bibitem{yates2021age}
R.~D. Yates, Y.~Sun, R.~D. Brown, S.~K. Kaul, E.~Modiano, and S.~Ulukus.
\newblock Age of information: An introduction and survey.
\newblock {\em IEEE J. Sel. Areas Commun.}, 39(5):1183--1210, May 2021.

\bibitem{kaul2012real}
S.~Kaul, R.~D. Yates, and M.~Gruteser.
\newblock Real-time status: How often should one update?
\newblock In {\em Proc. IEEE INFOCOM}, March 2012.

\bibitem{9516691}
A.~Arafa, J.~Yang, S.~Ulukus, and H.~V. Poor.
\newblock Timely status updating over erasure channels using an energy
  harvesting sensor: Single and multiple sources.
\newblock {\em IEEE Trans. Green Commun. Netw.}, 6(1):6--19, 2022.

\bibitem{yang2020age}
H.~Yang, A.~Arafa, T.~Q.~S Quek, and H.~V. Poor.
\newblock Age-based scheduling policy for federated learning in mobile edge
  networks.
\newblock In {\em Proc. IEEE ICASSP}, April 2020.

\bibitem{yates2021age_gossip}
R.~D. Yates.
\newblock The age of gossip in networks.
\newblock In {\em Proc. IEEE ISIT}, July 2021.

\bibitem{9589847}
J.~Heand, Q.~Ma, M.~Zhang, and J.~Huang.
\newblock Optimal fresh data sampling and trading.
\newblock In {\em Proc. IEEE WiOpt}, October 2021.

\bibitem{sac2018age}
H.~Sac, T.~Bacinoglu, E~Uysal-Biyikoglu, and G.~Durisi.
\newblock Age-optimal channel coding blocklength for an {M/G/1} queue with
  {HARQ}.
\newblock In {\em Proc. IEEE SPAWC}, June 2018.

\bibitem{abd2019role}
M.~A. Abd-Elmagid, N.~Pappas, and H.~S. Dhillon.
\newblock On the role of age of information in the internet of things.
\newblock {\em IEEE Commun. Mag.}, 57(12):72--77, December 2019.

\bibitem{zhao2024optimizing}
F.~Zhao, N.~Pappas, C.~Ma, X.~Sun, T.~Q.~S Quek, and H.~H Yang.
\newblock Optimizing information freshness in mobile networks with
  age-threshold aloha.
\newblock In {\em Proc. IEEE ISIT}, July 2024.

\bibitem{he2024age}
X.~He, C.~You, and T.~Q.~S. Quek.
\newblock Age-based scheduling for mobile edge computing: A deep reinforcement
  learning approach.
\newblock {\em IEEE Trans. Mobile Comput.}, 23(10):9881--9897, February 2024.

\bibitem{arafa2023private}
A.~Arafa and K.~Banawan.
\newblock Private status updating with erasures: A case for retransmission
  without resampling.
\newblock In {\em Proc. IEEE ICC}, May 2023.

\bibitem{minnaard2014timestomping}
W.~Minnaard.
\newblock Timestomping {NFTS}.
\newblock Master's thesis, University of Amsterdam, July 2014.

\bibitem{kaswan2024timestomping}
P.~Kaswan and S.~Ulukus.
\newblock Timestomping vulnerability of age-sensitive gossip networks.
\newblock {\em IEEE Trans. Commun.}, 72(7):4193--4205, July 2024.

\bibitem{ward2017use}
D.~Ward.
\newblock {\em Remote Estimation Over Use-Dependant Channels}.
\newblock PhD thesis, University of Maryland, 2017.

\bibitem{boyd2004convex}
S.~Boyd and L.~Vandenberghe.
\newblock {\em Convex optimization}.
\newblock Cambridge university press, 2004.

\bibitem{dinkelbach1967nonlinear}
W.~Dinkelbach.
\newblock "{O}n nonlinear fractional programming".
\newblock {\em Manag. Sci.}, 13(7):492--498, 1967.

\bibitem{Banawan_journal_2023}
K.~Banawan, A.~Arafa, and K.~G. Seddik.
\newblock Timely multi-process estimation over erasure channels with and
  without feedback: Signal-independent policies.
\newblock {\em IEEE J. Sel. Areas Inf. Theory}, 4:607--623, November 2023.

\end{thebibliography}

\end{document}